\documentclass[twocolumn,showpacs,showkeys,prb,aps]{revtex4}
\usepackage[writekey=false]{notes2bib}
 
\begin{document}
\preprint{revised preprint xxx.lanl.gov physics/9907051}
 
\title{Orthogonal Linear Combinations of Gaussian Type Orbitals}
 
\author{Richard J. Mathar}
\affiliation{
Leiden Observatory, P.O. Box 9513, 2300 RA Leiden, The Netherlands}
\homepage{http://www.strw.leidenuniv.nl/~mathar}
\email{mathar@strw.leidenuniv.nl}
 
\date{\today}
\widetext
 
\begin{abstract}
The set of Gaussian Type Orbitals $g(n_1,n_2,n_3)$
of order $(n+1)(n+2)/2$ and
of common $n\equiv n_1+n_2+n_3\le 7$, common center and exponential, is
customized to define a set of $2n+1$ linear combinations $t_{n,m}$ $(-n\le m\le n)$
such that each
$t_{n,m}$
depends on the azimuthal and polar angle of the spherical coordinate
system like the real or imaginary part of the associated Spherical Harmonic.
(Results cover both Hermite and Cartesian Gaussian Type Orbitals.)
Overlap, kinetic energy and Coulomb energy
matrix elements are presented for generalized
basis functions of the type
$r^st_{n,m}$
($s=0,2,4\ldots$).
In addition,
normalization integrals $\int |g(n_1,n_2,n_3)|d^3r$ are calculated up to
$n=7$ and normalization integrals $\int |r^st_{n,m}|d^3r$ up to $n=5$.
\end{abstract}
\maketitle
 
\section{Scope}
 
Gaussian Type Orbitals (GTO's) are widely used construction elements of
basis sets in quantum chemistry.
They became highly successful owing to
fairly simple analytical representations of key integrals\cite{Boys,Lindh,Klopper,Yarkony,Zivko}
that pay off in terms of speed when the matrix elements are
calculated. This work deals with the conversion of Cartesian or
Hermite GTO's
which have a product representation in Cartesian coordinates
into harmonic oscillator functions which have a 
product representation in spherical coordinates.
The latter are orthogonal with respect to two quantum indices ---
the total number of integrals to be calculated is reduced by the
overlap and kinetic energy integrals over products of orbitals with
the same center.

Sec.\ \ref{secTerm} introduces part of the notation.
Sec.\ \ref{secMain} lists those linear combinations of Gaussians
defined in Cartesian coordinates that obtain the orthogonal property
and are found to be the harmonic oscillator eigenfunctions. To extend
coverage
of the functional space, the oscillator functions are generalized in
Sec.\ \ref{secGen}. Section \ref{secIntgs} computes
two-center overlap, kinetic energy and Coulomb integrals of those
without recourse to the expansions in Cartesian coordinates.
Sec.\ \ref{sec:produ} tabulates some of their products in support
of 2-particle Coulomb Integrals.
Absolute norms of orbitals play some role in
Fitting Function techniques and are given in Sec.\ \ref{secAbs}
for the types of Gaussians discussed before.

\section{Terminology}\label{secTerm}
Let
\begin{equation}
g(n_1,n_2,n_3)\equiv \alpha^{n/2}H_{n_1}(\sqrt{\alpha}x)H_{n_2}
(\sqrt{\alpha}y)H_{n_3}(\sqrt{\alpha}z)e^{-\alpha r^2}
\label{gdef}
\end{equation}
be a primitive Hermite GTO (HGTO) with exponent $\alpha$ and quantum number $n\equiv n_1+n_2+n_3$
centered at the origin. $H_{n_i}$ are Hermite Polynomials, and $r^2\equiv
x^2+y^2+z^2$.
It is normalized as\cite{Zivko}\bibnote[doublefac]{
$k!!\equiv 1\cdot 3\cdot 5\cdots k$ for odd $k$.
$k!!\equiv 2\cdot 4\cdot 6\cdots k$ for even $k$.
$(-1)!!\equiv 0!!\equiv 1$.}
\begin{equation}
\int g^2(n_1,n_2,n_3)d^3r = \left(\frac{\pi}{2\alpha}\right)^{3/2}
\alpha^n\prod_{j=1}^3 (2n_j-1)!!.
\label{normg}
\end{equation}
$(n+1)(n+2)/2$ different $g(n_1,n_2,n_3)$ exist for a given $n$.
If $n\ge 2$, they build an overcomplete set of states compared to only $2n+1$
eigenstates $Y_n^m$ of the angular momentum operator.
\begin{equation}
Y_l^m(\theta,\varphi)\equiv
\left\{\frac{2l+1}{4\pi}\frac{(l-|m|)!}{(l+|m|)!}\right\}^{1/2}
P_l^{|m|}(\cos\theta)e^{im\varphi}
\end{equation}
($-l\le m\le l$) shall denote Spherical Harmonics in spherical
coordinates,\cite{Bradley1972}\bibnote[plnsign]
{Definitions with an additional sign $(-)^m$ or $(-)^{(m+|m|)/2}$ are also in frequent use.}
and
\begin{equation}
P_l^m(u)\equiv (1-u^2)^{m/2}\frac{d^m}{du^m}P_l(u)
\label{plmdef}
\end{equation}
generalized Legendre Polynomials.
Their real-valued counterparts are
\begin{eqnarray}
Y_l^{m,c}&\equiv& (Y_l^m+Y_l^{-m})/\sqrt{2}
\propto P_l^m(\cos\theta)\cos(m\varphi), \label{Ylmc} \\
Y_l^{m,s}&\equiv& -i(Y_l^m-Y_l^{-m})/\sqrt{2}
\propto P_l^m(\cos\theta)\sin(m\varphi), \label{Ylms}
\end{eqnarray}
and
\begin{equation}
Y_l^{0,c}\equiv Y_l^0 \propto P_l^0(\cos\theta),
\end{equation}
($0<m\le l$), all normalized to unity, with parity $(-)^l$, and
orthogonal,
\begin{eqnarray}
&& \int_0^{\pi} \sin\theta d\theta \int_0^{2\pi}d\varphi
Y_l^{m,c}(\theta,\varphi)Y_{l'}^{m',c}(\theta,\varphi) \nonumber \\
&&\quad =
\int_0^{\pi} \sin\theta d\theta \int_0^{2\pi}d\varphi
Y_l^{m,s}(\theta,\varphi)Y_{l'}^{m',s}(\theta,\varphi) \nonumber \\
&&\quad =\delta_{ll'}\delta_{mm'} ;\\
&& \int_0^{\pi} \sin\theta d\theta \int_0^{2\pi}d\varphi
Y_l^{m,c}(\theta,\varphi)Y_{l'}^{m',s}(\theta,\varphi)
= 0
.
\end{eqnarray}
The main result of this work is support to use of GTO's in systems with
heavy or
highly polarized atoms by reduction of the overcomplete sets to
sets of $2n+1$ linearly independent combinations of GTO's.
 
\section{Complete Sets of Gaussian Type Orbitals}\label{secMain}
\subsection{Hermite Basis}\label{secmain}
Below, HGTO's are linearly combined into sets of real-valued functions
$t_{n,m}(r,\theta,\varphi,\alpha)$ ($-n\le m\le n$) that
display angular dependencies $t_{n,m}\propto Y_n^{m,c}(\theta,\varphi)$
for $m\ge 0$ and
$t_{n,m}\propto Y_n^{-m,s}(\theta,\varphi)$ for $m< 0$ by way of
construction (see Appendix \ref{appA})\@.
Hence they are orthogonal\bibnote{
If orthogonality of the basis vectors is unimportant, other, much more
general sets of linear combinations with fewer $g$-terms
might help to reduce the total amount of time spent in numerical integrals
calculations.
(J. C. Boetter, priv.\ commun.)
}
\begin{equation}
\int_0^{\pi} \sin\theta d\theta \int_0^{2\pi}d\varphi t_{n,m}t_{n',m'}
\propto \delta_{n,n'}\delta_{m,m'}
\end{equation}
and complete with respect to the angular variables.
Their norms are listed in form of
\begin{equation}
\int t_{n,m}^2 d^3r=N_{nm}\sqrt{2\pi^3 \alpha^{-3}}\alpha^n.
\label{normt}
\end{equation}
The $N_{nm}$ follow each time after three dots and are easily derived from
the expansion coefficients given and the overlap integrals\citenote{doublefac}
\bibnote[Gaussbra]{$\lfloor x\rfloor$ denotes the largest integer $\leq x$.}
\begin{eqnarray}
&& \int g(n_1,n_2,n_3)g(n_1',n_2',n_3')d^3r \nonumber \\
&& =\Biggm\{
\begin{array}{r}
\displaystyle \alpha^{(n+n')/2}\left(\frac{\pi}{2\alpha}\right)^{3/2}
\prod_{j=1}^3
(-)^{\lfloor\frac{\scriptstyle n_j}{\scriptstyle 2}\rfloor+\lfloor\frac{\scriptstyle n_j'}{\scriptstyle 2}\rfloor}
(n_j+n_j'-1)!!  \\
,\text{all}\, n_j+n_j'\, \text{even}\\
0 \qquad , \text{any}\, n_j+n_j'\,  \text{odd}\label{goverl}\\
\end{array}
\end{eqnarray}
using Eq.\ (7.374.2) of Ref.\ \onlinecite{GR} or
Eqs.\ (20) and (21) of Ref.\ \onlinecite{Apelblat}\@.
The radial dependence $t_{n,m}\propto (\alpha r)^n\exp(-\alpha r^2)$ is found by inspection
of the integrals mentioned in item 1 of App.~\ref{appA}, which leads to
\begin{equation}
t_{n,m}=
2^{n+2}\sqrt{\frac{\pi N_{n,m}}{(2n+1)!!}}
(\alpha r)^ne^{-\alpha r^2}Y_n^{m\{c,s\}} ,
\label{tnmdef}
\end{equation}
where
\begin{equation}
Y_n^{m\{c,s\}}\equiv
\left\{
\begin{array}{cc} Y_n^{m,c}, & m\ge 0 \\ Y_n^{-m,s}, & m<0 
\end{array} .
\right.
\end{equation}
 
\begin{eqnarray*}
&t_{0,0}=g(0,0,0)\ldots 1/4 ; & \\
&t_{1,0}=g(0,0,1)\ldots 1/4 ; & \\
&t_{1,1}=g(1,0,0)\ldots 1/4 ; &\\
&t_{2,0}=2g(0,0,2)-g(2,0,0)-g(0,2,0)\ldots 3 ; &\\
&t_{2,1}=g(1,0,1)\ldots 1/4 ; &\\
&t_{2,2}=g(2,0,0)-g(0,2,0)\ldots 1 ; &\\
&t_{2,-2}=g(1,1,0)\ldots 1/4 ; &\\
&t_{3,0}=2g(0,0,3)-3(g(0,2,1)+g(2,0,1))\ldots 15 ;& \\
&t_{3,1}=4g(1,0,2)-g(3,0,0)-g(1,2,0)\ldots 10 ;& \\
&t_{3,2}=g(2,0,1)-g(0,2,1)\ldots 1 ;& \\
&t_{3,-2}=g(1,1,1)\ldots 1/4 ;& \\
&t_{3,3}=g(3,0,0)-3g(1,2,0)\ldots 6 ;& \\
&t_{4,0}=8g(0,0,4)+6g(2,2,0)-24(g(2,0,2)+g(0,2,2))& \\
   &+3(g(4,0,0)+g(0,4,0))\ldots 1680 ;& \\
&t_{4,1}=4g(1,0,3)-3g(1,2,1)-3g(3,0,1) \ldots 42 ;& \\
&t_{4,2}=6(g(2,0,2)-g(0,2,2))&\\
 &-g(4,0,0)+g(0,4,0) \ldots 84 ;& \\
&t_{4,-2}=6g(1,1,2)-g(1,3,0)-g(3,1,0) \ldots 21 ;& \\
&t_{4,3}=g(3,0,1)-3g(1,2,1) \ldots 6 ;& \\
&t_{4,4}=g(4,0,0)+g(0,4,0)-6g(2,2,0) \ldots 48 ;& \\
&t_{4,-4}=g(3,1,0)-g(1,3,0) \ldots 3 ;& \\
&t_{5,0}=8g(0,0,5)+15(g(4,0,1)+g(0,4,1)) &\\
   &-40(g(2,0,3)+g(0,2,3))+30g(2,2,1)\ldots 15120 ; &\\
&t_{5,1}=g(5,0,0)+2g(3,2,0)-12g(3,0,2)+g(1,4,0) &\\
   &+8g(1,0,4)-12g(1,2,2)\ldots 1008; & \\
&t_{5,2}=2(g(2,0,3)-g(0,2,3)) & \\
   &-g(4,0,1)+g(0,4,1)\ldots 36 ; & \\
&t_{5,-2}=2g(1,1,3)-g(3,1,1)-g(1,3,1)\ldots 9 ; & \\
&t_{5,3}=8g(3,0,2)-g(5,0,0)+2g(3,2,0) &\\
   &+3g(1,4,0)-24g(1,2,2)\ldots 864 ; & \\
&t_{5,4}=g(4,0,1)+g(0,4,1)-6g(2,2,1)\ldots 48 ; & \\
&t_{5,-4}=g(3,1,1)-g(1,3,1)\ldots 3; & \\
&t_{5,5}=g(5,0,0)-10g(3,2,0)+5g(1,4,0)\ldots 480; & \\
&t_{6,0}=16g(0,0,6)-5(g(0,6,0)+g(6,0,0)) &\\
  &-120(g(0,2,4)+g(2,0,4))+90(g(0,4,2)+g(4,0,2)) &\\
  &+180g(2,2,2) -15(g(2,4,0)+g(4,2,0))\ldots 665280 ; & \\
&t_{6,1}=8g(1,0,5)-20g(1,2,3)+5g(1,4,1)&\\
 &-20g(3,0,3) +10g(3,2,1)+5g(5,0,1)\ldots 7920 ; & \\
&t_{6,2}=g(6,0,0)-g(0,6,0)+g(4,2,0)\hfil&\\
 &-g(2,4,0) +16(g(0,4,2)-g(4,0,2))&\\
 &+16(g(2,0,4)-g(0,2,4))\ldots 12672; & \\
&t_{6,-2}=16g(1,1,4)-16(g(1,3,2)+g(3,1,2)) &\\
&+g(1,5,0)+g(5,1,0)+2g(3,3,0)\ldots 3168; & \\
&t_{6,3}=8g(3,0,3)-24g(1,2,3)+9g(1,4,1) &\\
   &+6g(3,2,1)-3g(5,0,1)\ldots 3168; &\\
&t_{6,4}=10(g(0,4,2)+g(4,0,2))&\\
 &+5(g(2,4,0)+g(4,2,0))-60g(2,2,2)&\\
 & -g(0,6,0)-g(6,0,0)\ldots 10560; &\\
&t_{6,-4}=g(1,5,0)-g(5,1,0)&\\
 &+10(g(3,1,2)-g(1,3,2))\ldots 660; &\\
&t_{6,5}=g(5,0,1)+5g(1,4,1)-10g(3,2,1)\ldots 480; &\\
&t_{6,6}=g(6,0,0)-g(0,6,0)&\\
 &+15(g(2,4,0)-g(4,2,0))\ldots 5760; &\\
&t_{6,-6}=3(g(1,5,0)+g(5,1,0))-10g(3,3,0)\ldots 1440; &\\
&t_{7,0}= 16g(0,0,7)-35(g(0,6,1)+g(6,0,1))&\\
  &-105(g(2,4,1)+g(4,2,1))-168(g(0,2,5)+g(2,0,5)) &\\
  &+210(g(0,4,3)+g(4,0,3))+420g(2,2,3)
\ldots 8648640 ; &\\
&t_{7,1}=240g(3,2,2)-5g(1,6,0)+64g(1,0,6)&\\
 &-240g(1,2,4) +120g(1,4,2)-240g(3,0,4)-5g(7,0,0)&\\
 &-15g(3,4,0) +120g(5,0,2)-15g(5,2,0) \ldots 4942080; &\\
&t_{7,2}=15(g(6,0,1)-g(0,6,1))&\\
 &+15(g(4,2,1)-g(2,4,1)) +48(g(2,0,5)-g(0,2,5))&\\
 &+80(g(0,4,3)-g(4,0,3))\ldots 823680; &\\
&t_{7,-2}=48g(1,1,5)-80(g(1,3,3)+g(3,1,3))&\\
  &+15(g(1,5,1)+g(5,1,1))+30g(3,3,1)\ldots 205920; &\\
&t_{7,3}=3g(7,0,0)-240g(1,2,4)+180g(1,4,2)&\\
 &-9g(1,6,0)+80g(3,0,4) +120g(3,2,2)&\\
 &-15g(3,4,0)-60g(5,0,2)-3g(5,2,0)\ldots 1647360; &\\
&t_{7,4}=10(g(0,4,3)+g(4,0,3))-3(g(6,0,1)+g(0,6,1))&\\
  &-60g(2,2,3)+15(g(2,4,1)-g(4,2,1))\ldots 37440; &\\
&t_{7,-4}=3(g(1,5,1)-g(5,1,1))&\\
 &+10(g(3,1,3)-g(1,3,3))\ldots 2340 ;&\\
&t_{7,5}=60g(1,4,2)-g(7,0,0)-5g(1,6,0)-120g(3,2,2)&\\
  &+5g(3,4,0) +12g(5,0,2)+9g(5,2,0)\ldots 149760; &\\
&t_{7,6}=g(6,0,1)-g(0,6,1)&\\
  &+15(g(2,4,1)-g(4,2,1))\ldots 5760 ;&\\
&t_{7,-6}=3(g(1,5,1)+g(5,1,1))-10g(3,3,1)\ldots 1440 ;&\\
&t_{7,7}=g(7,0,0)-7g(1,6,0)+35g(3,4,0)&\\
 &-21g(5,2,0)\ldots 80640 ; & \\
&t_{8,0}=128g(0,0,8)+35(g(0,8,0)+g(8,0,0)) &\\
  &+3360(g(0,4,4)+g(4,0,4)) -1792(g(0,2,6)+g(2,0,6))&\\
 &-1120(g(0,6,2)+g(6,0,2))+6720g(2,2,4)&\\
  &-3360(g(2,4,2)+g(4,2,2))+140(g(2,6,0)+g(6,2,0))& \\
& +210(4,4,0) \ldots 8302694400 ; &\\
&t_{8,8}=g(8,0,0)+g(0,8,0)-28(g(6,2,0)+g(2,6,0))&\\
&+70g(4,4,0)\ldots 1290240 ; &\\
&t_{8,-8}=g(7,1,0)-g(1,7,0) & \\
& +7(g(3,5,0)-g(5,3,0))\ldots 20160
.&
\end{eqnarray*}
$t_{n,-m}$ with odd $m$ are not shown explicitly, but
are incorporated implicitly by an interchange of the first two
arguments of
every $g$ on the right hand side of $t_{n,m}$ and multiplication
by $(-1)^{\lfloor m/2\rfloor}$, like for example
\begin{equation}
t_{4,-1}=4g(0,1,3)-3g(2,1,1)-3g(0,3,1) \ldots 42
.
\end{equation}
This follows from applying the mirror operation $x\leftrightarrow y$
to the equations shown, which is
$\varphi \leftrightarrow \pi/2-\varphi$
in polar coordinates, and induces
$\cos(m\varphi)\leftrightarrow (-1)^{\lfloor m/2\rfloor}\sin(m\varphi)$ and
$Y_n^{m,c}\leftrightarrow (-1)^{\lfloor m/2\rfloor} Y_n^{m,s}$
if $m$ is odd.
 
Application of the Laplace operator yields
\begin{equation}
\nabla^2 t_{n,m}=2\alpha (2\alpha r^2-2n-3)t_{n,m}.
\label{lapla}
\end{equation}
The $t_{n,m}$ with exponential $\alpha =M\omega/(2\hbar)$ are eigenfunctions of the
3-dimensional isotropic harmonic oscillator
\begin{equation}
(-\frac{\hbar^2}{2M}\nabla^2+\frac{1}{2}M\omega^2r^2)t_{n,m}
=(n+\frac{3}{2})\hbar\omega t_{n,m}.
\end{equation}

\subsection{Cartesian Basis}
 
The previous list contains also the expansions in terms of
Cartesian GTO's (CGTO's) as outlined in App.~\ref{appB}\@.
If each $g(n_1,n_2,n_3)$ in the list is replaced by the
primitive CGTO of the same triple index,
\begin{equation}
f(n_1,n_2,n_3)\equiv x^{n_1}y^{n_2}z^{n_3}e^{-\alpha r^2},
\end{equation}
and each $t_{n,m}$ by $\tilde t_{n,m}$,
the newly defined $\tilde t_{n,m}$ are $\propto r^n\exp(-\alpha r^2)Y_n^{m,c}$
($m\ge 0$) and $\propto r^n\exp(-\alpha r^2)Y_n^{-m,s}$ ($m<0$),
for example
\begin{eqnarray}
\tilde t_{4,1}&=&4f(1,0,3)-3f(1,2,1)-3f(3,0,1) \nonumber \\
&=& \frac{4}{3}\sqrt{\frac{2\pi}{5}}r^4e^{-\alpha r^2}Y_4^{1,c}(\theta,\varphi) .
\end{eqnarray}
The normalization integrals
are recovered from the Kronecker product
of the vectors of the expansion coefficients and the overlap
integrals\citenote{doublefac}
\begin{eqnarray}
&& \int f(n_1,n_2,n_3)f(n_1',n_2',n_3')d^3r \nonumber \\
&& =\Biggm\{
\begin{array}{r}
\displaystyle \frac{1}{(4\alpha )^{(n+n')/2}}\left(\frac{\pi}{2\alpha }\right)^{3/2}
\prod_{j=1}^3 (n_j+n_j'-1)!! \\
,\text{all}\, n_j+n_j'\, \text{even}\\
0 \qquad , \text{any}\, n_j+n_j' \,\text{odd}. \label{foverl} \\
\end{array}
\end{eqnarray}
to become
\begin{equation}
\int \tilde t^2_{n,m}d^3r = N_{nm}\sqrt{2\pi^3\alpha ^{-3}}/(4\alpha )^n .
\label{normttilde}
\end{equation}
[The similarity between Eq.\ (\ref{goverl}) and Eq.\ (\ref{foverl}) lets the
$N_{nm}$ already given in Eq.\ (\ref{normt}) show up again.
The additional
sign in Eq.\ (\ref{goverl}) is positive if applied to $g$-terms
of a single $t_{n,m}$, and does not mix things up.]

A synopsis of Eqs.\ (\ref{normt}), (\ref{tnmdef}) and (\ref{normttilde})
demonstrates
\begin{equation}
\tilde t_{n,m}=t_{n,m}/(2\alpha)^n .
\end{equation}

An explicit expansion of the vector harmonics multiplied with the Gaussian
in terms of CGTO's is derived in App.~\ref{appX}:\citenote{Gaussbra}
\begin{eqnarray}
&& r^ne^{-\alpha r^2}Y_n^{m\{c,s\}}=
\frac{1}{2^m}
\sqrt{\frac{2n+1}{2\pi}(n-|m|)!(n+|m|)!} \nonumber \\
&\times&
\sum_{\sigma_1,\sigma_2\ge 0}^{\sigma_1+\sigma_2\le \lfloor (n-|m|)/2\rfloor}
\frac{1}{\sigma_1!\sigma_2!}\left(-\frac{1}{4}\right)^{\sigma_1+\sigma_2}
\frac{1}{(|m|+\sigma_1+\sigma_2)!}
\nonumber \\
&\times&
\frac{1}{(n-|m|-2(\sigma_1+\sigma_2))!}
\sum_{j=0(1)}^{|m|}(-)^{\lfloor j/2\rfloor}
{|m| \choose j}
\nonumber \\
&\times&
f(|m|-j+2\sigma_1,j+2\sigma_2,n-|m|-2(\sigma_1+\sigma_2)).
\label{fasttnm2}
\end{eqnarray}
(In case of $Y_n^{m,c}$ the sum over $j$ attains only even values of $j$,
in case of $Y_n^{m,s}$ only odd values.)
This triple sum is $\tilde t_{n,\pm m}$
up to a normalization factor that depends on $n$ and $m$,
and therefore (up to a different factor)
also $t_{n,\pm m}$ when $f$ is replaced by $g$.
 
\subsection{Rayleigh-Type Expansion}
The expansion of a plane wave in terms of the
$t_{n,m}$ reads
\begin{eqnarray}
e^{i{\bf k}\cdot{\bf r}}=&&2e^{-k^2/(4\alpha )}
\sum_{n=0}^{\infty}
\left(\frac{ik}{\alpha}\right)^n \nonumber \\
&\times&\sum_{m=-n}^n\sqrt{\frac{2\pi}{(2n+1)!!N_{nm}}}
Y_n^{m\{c,s\}}(\widehat{\bf k})t_{n,m}
,
\end{eqnarray}
similar to the Rayleigh expansion
\begin{equation}
e^{i{\bf k}\cdot{\bf r}}
=4\pi\sum_{l=0}^{\infty}\sum_{m=-l}^l
i^lj_l(kr)Y_l^{m*}(\hat{\bf k})Y_l^m(\hat{\bf r}), \label{Rayl}
\end{equation}
where
$\widehat{\bf r}$ denotes the angular variables of the vector $\bf r$.
The expansion coefficients are the quotients of the integrals
$\langle e^{i{\bf k}\cdot{\bf r}}
|t_{n,m}\rangle/\langle t_{n,m}|t_{n,m}\rangle$.
If ${\bf r}$ points into the $z$-direction, we have
\begin{equation}
Y_n^{m\{c,s\}}(\theta_k=0,\varphi_k)=\sqrt{\frac{2n+1}{4\pi}}\delta_{m,0},
\end{equation}
whence
\begin{eqnarray}
&& e^{ikz}=\sqrt{2}e^{-k^2/(4\alpha )}\Bigg\{2t_{0,0}
+2i\frac{k}{\alpha }t_{1,0}
-\frac{1}{3}\left(\frac{k}{\alpha }\right)^2 t_{2,0} \nonumber \\
&& -\frac{i}{15}\left(\frac{k}{\alpha }\right)^3 t_{3,0}
+\frac{1}{420}\left(\frac{k}{\alpha }\right)^4 t_{4,0}
+\frac{i}{3780}\left(\frac{k}{\alpha }\right)^5 t_{5,0}\cdots\Bigg\}
\nonumber \\
&& =\sqrt{2}e^{-k^2/(4\alpha)}
\sum_{n=0}^{\infty}\frac{1}{\sqrt{(2n-1)!!N_{n0}}}
\left(\frac{ik}{\alpha}\right)^nt_{n,0}
. \label{rayl}
\end{eqnarray}
The closely related Cartesian case is written down by  replacing
$t_{n,0}$ with $(2\alpha )^n \tilde t_{n,0}$ in Eq.\ (\ref{rayl})
for all $n$.

\section{Generalized Oscillator Functions}\label{secGen}
\subsection{Definition}

The basis of the $t_{n,m}$
defined in the previous sections
is incomplete with respect to the radial coordinate.
The combination $g(2,0,0)+g(0,2,0)+g(0,0,2)$ of HGTO's
contains a component $\propto r^2e^{-\alpha r^2}Y_0^{0,c}$, for example,
which cannot be represented
in terms of $t_{n,m}$. The subsequent sections therefore introduce
a more general type of functions centered at space points 
${\bf A}$, which is derived from Eq.\ (\ref{tnmdef}) by multiplication with $r^s$:
\begin{eqnarray}
t_{n,m,s}(\alpha,{\bf r-A})&\equiv&
2^{n+2}\sqrt{\frac{\pi N_{n,m}}{(2n+1)!!}}
|{\bf r-A}|^s(\alpha |{\bf r-A}|)^n \nonumber \\
&&\times e^{-\alpha ({\bf r-A})^2}
Y_n^{m\{c,s\}}(\widehat{\bf r-A}) .
\label{tnmsdef}
\end{eqnarray}

All subsequent considerations assume that $s$ is a non-negative even integer.
(i)~This ensures that $t_{n,m,s}$ is another linear combination of HGTO's,
which is made explicit in App.~\ref{appC}\@. Therefore a standard, indirect path of computing
integrals is already established: linkage to known approaches\cite{Boys,Lindh,Klopper,Yarkony} by dissociation
of all $t_{n,m,s}$ in the integrands.
A relevant application is given in Sec.\ \ref{sec:produ}.
(ii)~The restriction leads to truncation of some series expressions that follow ---
important to numerical evaluation --- whereas
odd $s$ would sometimes yield infinite series.
(iii)~$s+n>-3/2$ is needed to guarantee convergence of the normalization integral,
and (iv)~the set of $s=0,2,4,\ldots$ suffices to let the $t_{n,m,s}$ span the
entire vector space of the GTO's.

The generalization of Eq.\ (\ref{normt}) reads
\begin{equation}
\int t_{n,m,s}^2 d^3r=N_{nm}\frac{(2s+2n+1)!!}{(2n+1)!!(4\alpha)^s}
\sqrt{2\pi^3 \alpha^{-3}}\alpha^n,
\label{tnmnorm}
\end{equation}
and the generalization of Eq.\ (\ref{lapla})
\begin{eqnarray}
\nabla^2 t_{n,m,s}(\alpha,{\bf r})
&=&\Big[s(2n+s+1)\frac{1}{r^2}
 -4\alpha \left(s+n+\frac{3}{2}\right)
\nonumber \\
&&
+4\alpha^2 r^2\Big]
t_{n,m,s}(\alpha,{\bf r}).
\label{lapla2}
\end{eqnarray}

$t_{n,m,s}$ is an eigenfunction of a spherical potential
with repulsive core as detailed in App.~\ref{appD}.

\subsection{Fourier (Momentum) Representation}
The Fourier integral of the orbital centered at the origin
is related to Eq.\ (\ref{fourrep}).
The Rayleigh expansion (\ref{Rayl}) is inserted,
then the radial integral is solved and re-written with Eqs.\ (11.4.28) 
and (13.1.27) of Ref.\ \onlinecite{AS}:
\begin{eqnarray}
&&t_{n,m,s}(\alpha,{\bf k})\equiv
\int e^{i{\bf k}\cdot{\bf r}}t_{n,m,s}(\alpha,{\bf r})d^3r
\nonumber \\
&=&
4\pi^2\sqrt{\frac{N_{n,m}}{(2n+1)!!}}
(ik)^n\frac{\Gamma(n+\frac{3}{2}+\frac{s}{2})}{\Gamma(n+\frac{3}{2})\alpha^{(3+s)/2}}
\nonumber \\
&&\times
e^{-k^2/(4\alpha)}{}\underbrace{_1F_1(-\frac{s}{2};n+\frac{3}{2};\frac{k^2}{4\alpha})}
_{\displaystyle\sum_{\sigma=0}^{s/2}\frac{(-s/2)_{\sigma}}{\sigma !(n+3/2)_{\sigma}}
\left(\frac{k^2}{4\alpha}\right)^{\sigma}}
Y_n^{m\{c,s\}}(\widehat{\bf k}) ,
\label{four2} \\
&=&
4\pi^2\sqrt{\frac{N_{n,m}}{(2n+1)!!}}
(ik)^n\frac{(s/2)!}{\alpha^{(3+s)/2}}
\nonumber \\
&&\times
e^{-k^2/(4\alpha)}L_{s/2}^{(n+1/2)}\left(\frac{k^2}{4\alpha}\right)
Y_n^{m\{c,s\}}(\widehat{\bf k})
, \nonumber
\end{eqnarray}
where Pochhammer's Symbol\bibnote[Poch]{$(x)_p\equiv x\cdot (x+1)\cdot(x+2)\cdots (x+p-1)$
for positive integers $p$.
$(x)_0\equiv 1$.}
has been used.

\section{Integrals of Oscillator Functions}\label{secIntgs}
\subsection{Overlap Integral}
The overlap integral is a convolution in real space and calculated
with the shift theorem of  Fourier analysis by use of Eq.\ (\ref{four2}) with $t_{n,m,s}(\alpha,{\bf k})$
and $t_{n',m',s'}(\beta,-{\bf k})$:
\begin{eqnarray}
&&\int t_{n,m,s}(\alpha,{\bf r-A})t_{n',m',s'}(\beta,{\bf r-B}) d^3r\nonumber \\
&=&
\int t_{n,m,s}(\alpha,{\bf r})t_{n',m',s'}(\beta,{\bf r-C})d^3r \nonumber \\
&=&
\int \frac{d^3k}{(2\pi)^3}e^{-i{\bf k}\cdot{\bf C}} t_{n,m,s}(\alpha,{\bf k})
t_{n',m',s'}(\beta,-{\bf k})\nonumber \\
&=&
2^{n+n'}(4\pi)^2\sqrt{\frac{\pi N_{n,m}N_{n',m'}}{(2n+1)!!(2n'+1)!!\alpha^{s+3}\beta^{s'+3}}}
e^{-\gamma C^2}
\nonumber \\
&\times&
\gamma^{(n+n'+3)/2}
\Gamma\left(n+\frac{3}{2}+\frac{s}{2}\right)
\Gamma\left(n'+\frac{3}{2}+\frac{s'}{2}\right)
\nonumber \\
&\times&
\!\sum_{\sigma=0}^{s/2}\frac{(-s/2)_\sigma}{\sigma !\Gamma(n+\frac{3}{2}+\sigma)}\left(\frac{\gamma}{\alpha}\right)^\sigma
\sum_{\sigma'=0}^{s'/2}\frac{(-s'/2)_{\sigma'}}{\sigma' !\Gamma(n'+\frac{3}{2}+\sigma')}\left(\frac{\gamma}{\beta}\right)^{\sigma'}
\nonumber \\
&\times&
\!\!\!\sum_{l=|n-n'|}^{n+n'}
(-)^{(n-n'-l)/2}
(\gamma C^2)^{l/2}
\Gamma(\frac{3}{2}+\frac{n+n'+l}{2}+\sigma+\sigma')
\nonumber \\
&\times&
y_{lnn'}^{mm'}(\hat{\bf C})
\sum_{\tilde\sigma=0}^{(n+n'-l)/2+\sigma+\sigma'}
\frac{(\frac{l-n-n'}{2}-\sigma-\sigma')_{\tilde\sigma}}{\tilde\sigma !
\Gamma(l+\frac{3}{2}+\tilde\sigma)}(\gamma C^2)^{\tilde \sigma}. \nonumber \\
&&
\label{overl2}
\end{eqnarray}
Here ${\bf C}\equiv {\bf B-A}$ and $\gamma\equiv \alpha\beta/(\alpha+\beta)$.
The angular integral $y_{lnn'}^{mm'}$ is expressed in terms of
Wigner's $3j$-Symbol:\cite{Edmonds}\cite[\S $3^3$]{Condon}\cite[Chap.\ XIV]{LandauLifshitz3}
\begin{eqnarray*}
&&y_{lnn'}^{mm'}(\hat{\bf C})\equiv
\sum_{\kappa=-l}^l
Y_l^{\kappa *}(\hat{\bf C})\int\sin\theta_k d\theta d\varphi_k \nonumber \\
&&\qquad\qquad \times
Y_l^{\kappa}(\hat{\bf k})Y_n^{m\{c,s\}}(\hat{\bf k})
Y_{n'}^{m'\{c,s\}}(\hat{\bf k})
\nonumber \\
&&=
\sqrt{\frac{(2l+1)(2n+1)(2n'+1)}{4\pi}}
\left(
\begin{array}{ccc} l& n& n' \\ 0 & 0 & 0
\end{array}
\right)\bar y_{lnn'}^{mm'}(\hat{\bf C}).
\end{eqnarray*}
Eq.\ (4.6.3) of Ref.\ \onlinecite{Edmonds} and adaptation to its sign convention
of Spherical Harmonics determine $\bar y_{lnn'}^{mm'}(\hat{\bf C})$:
\begin{eqnarray}
&&m'=0: \nonumber \\
&&\bar y_{lnn'}^{mm'}
= (-)^{m}\left( \begin{array}{ccc} l & n & n' \\ m & -m & 0 \end{array}\right)
	Y_l^{m\{c,s\}}(\hat{\bf C}); \label{firsty}
\\
&&m>m'>0:\nonumber \\
&& \bar y_{lnn'}^{mm'}
= \frac{(-)^{m+m'}}{\sqrt{2}}
   \left( \begin{array}{ccc} l & n & n' \\ m+m' & -m & -m' \end{array}\right)
	Y_l^{m+m',c}(\hat{\bf C}) \nonumber \\
     &&\quad +\frac{(-)^m}{\sqrt{2}}
   \left( \begin{array}{ccc} l & n & n' \\ m-m' & -m & m' \end{array}\right)
	Y_l^{m-m',c}(\hat{\bf C})
      ;
\\
&&m=m'>0:\nonumber \\
&& \bar y_{lnn'}^{mm'}
 = \frac{1}{\sqrt{2}}
   \left( \begin{array}{ccc} l & n & n' \\ m+m' & -m & -m' \end{array}\right)
	Y_l^{m+m',c}(\hat{\bf C}) \nonumber \\
&&\quad     +(-)^m
   \left( \begin{array}{ccc} l & n & n' \\ 0 & -m & m' \end{array}\right)
	Y_l^{0,c}(\hat{\bf C})
      ;
\\
&&m<m'<0: \nonumber \\
&&\bar y_{lnn'}^{mm'}
 = -\frac{(-)^{m+m'}}{\sqrt{2}}
   \left( \begin{array}{ccc} l & n & n' \\ m+m' & -m & -m' \end{array}\right)
	Y_l^{|m+m'|,c}(\hat{\bf C}) \nonumber \\
&&\quad     +\frac{(-)^m}{\sqrt{2}}
   \left( \begin{array}{ccc} l & n & n' \\ m-m' & -m & m' \end{array}\right)
	Y_l^{|m-m'|,c}(\hat{\bf C})
      ;
\\
&&m=m'<0: \nonumber \\
&&\bar y_{lnn'}^{mm'}
 = -\frac{1}{\sqrt{2}}
   \left( \begin{array}{ccc} l & n & n' \\ m+m' & -m & -m' \end{array}\right)
	Y_l^{|m+m'|,c}(\hat{\bf C}) \nonumber \\
&&\quad     +(-)^m
   \left( \begin{array}{ccc} l & n & n' \\ 0 & -m & m' \end{array}\right)
	Y_l^{0,c}(\hat{\bf C})
      ;
\\
&&m>|m'|,m'<0: \nonumber \\
&&\bar y_{lnn'}^{mm'}
 = \frac{(-)^{m+m'}}{\sqrt{2}}
   \left( \begin{array}{ccc} l & n & n' \\ m+|m'| & -m & -|m'| \end{array}\right)
	Y_l^{m+|m'|,s}(\hat{\bf C}) \nonumber \\
&&\quad     -\frac{(-)^m}{\sqrt{2}}
   \left( \begin{array}{ccc} l & n & n' \\ m-|m'| & -m & |m'| \end{array}\right)
	Y_l^{m-|m'|,s}(\hat{\bf C})
      ;
\\
&&0<m\le |m'|,m'<0: \nonumber \\
&&\bar y_{lnn'}^{mm'}
 = \frac{(-)^{m+m'}}{\sqrt{2}}
   \left( \begin{array}{ccc} l & n & n' \\ m+|m'| & -m & -|m'| \end{array}\right)
	Y_l^{m+|m'|,s}(\hat{\bf C}) \nonumber \\
&&\quad     +\frac{(-)^{m'}}{\sqrt{2}}
   \left( \begin{array}{ccc} l & n & n' \\ m-|m'| & -m & |m'| \end{array}\right)
	Y_l^{-m+|m'|,s}(\hat{\bf C})
      .
   \label{lasty}
\end{eqnarray}
[Occurrences of $Y_n^{0,s}$ have to be interpreted as $0$ in Eq.\ (\ref{lasty}).]
Some cases like $m<0, m'>0$ or $m'<m<0$ are not covered
by Eqs.\ (\ref{firsty})--(\ref{lasty}), but may be dispatched by help
of the permutation symmetries
\begin{equation}
y_{lnn'}^{mm'}(\hat{\bf C})=
y_{ln'n}^{m'm}(\hat{\bf C})\quad ; \quad
\bar y_{lnn'}^{mm'}(\hat{\bf C})=
\bar y_{ln'n}^{m'm}(\hat{\bf C}).
\end{equation}

In case of common center (${\bf C}={\bf 0}$),
the orthogonality relation\cite[\S (4b)]{Mitroy}
\begin{eqnarray}
&&\int t_{n,m,s}(\alpha,{\bf r-A})t_{n',m',s'}(\beta,{\bf r-A})d^3r \nonumber \\
&&= \delta_{n,n'}\delta_{m,m'}\frac{2^{2n+3}\pi N_{n,m}}{(2n+1)!!}\gamma^n
\frac{\Gamma(n+\frac{3}{2}+\frac{s}{2}+\frac{s'}{2})}{(\alpha+\beta)
^{(s+s'+3)/2}}
\label{tortho}
\end{eqnarray}
results, which turns into Eq.\ (\ref{tnmnorm}) if $\alpha=\beta$ and $s=s'$.
 [If the derivation starts from Eq.\ (\ref{overl2}), the term $\sum_{\tilde\sigma}$ becomes
 $1/\Gamma(l+3/2)$. The factor $(\gamma c^2)^{l/2}$ then means that nonzero overlap
 can occur only if the term $l=0$ is present in the sum over $l$. In turn, $l=0$ becomes
 the only term that contributes, and inspection of the lower limit $l=|n-n'|$ introduces
 a factor $\delta_{nn'}$.
 Then $y_{0nn}^{mm'}=\delta_{mm'}/(4\pi)$, and eventually
 \begin{eqnarray}
 &&\sum_{\sigma=0}^{s/2}\frac{(-s/2)_{\sigma}}{\sigma !\Gamma(n+\frac{3}{2}+\sigma)}
 \left(\frac{\gamma}{\alpha}\right)^{\sigma}
 \sum_{\sigma'=0}^{s'/2}\frac{(-s'/2)_{\sigma'}}{\sigma' !\Gamma(n+\frac{3}{2}+\sigma')}
 \left(\frac{\gamma}{\beta}\right)^{\sigma'} \nonumber \\
 &&\times \Gamma(n+\frac{3}{2}+\sigma+\sigma')  \nonumber \\
 &&=\frac{1}{\Gamma(n+\frac{3}{2})}F_2(n+\frac{3}{2},-\frac{s}{2},-\frac{s'}{2},n+\frac{3}{2},
 n+\frac{3}{2};\frac{\gamma}{\alpha},\frac{\gamma}{\beta}) \nonumber \\
 \nonumber \\
 &&
 =
 \left(\frac{\gamma}{\beta}\right)^{s/2}
 \left(\frac{\gamma}{\alpha}\right)^{s'/2}
 \frac{\Gamma(n+\frac{3}{2}+\frac{s}{2}+\frac{s'}{2})}
 {\Gamma(n+\frac{3}{2}+\frac{s}{2})\Gamma(n+\frac{3}{2}+\frac{s'}{2})}
 \end{eqnarray}
 is calculated with Eqs.\ (9.180.2), (9.182.3) and (9.122.1) of Ref.\ \onlinecite{GR}.]
 
\subsection{Kinetic Energy Integral}
Two ``direct'' methods exist to calculate the kinetic energy matrix.
(i) Application of the kinetic energy operator on $t_{n',m',s'}(\beta,{\bf r-B})$ yields three
terms $\propto t_{n',m',s'-2}(\beta,{\bf r-B})$, $\propto t_{n',m',s'}(\beta,{\bf r-B})$
and $\propto t_{n',m',s'+2}(\beta,{\bf r-B})$
by Eq.\ (\ref{lapla2}) if $s'\neq 0$, and two terms by Eq.\ (\ref{lapla}) if $s'=0$.
Kinetic energy integrals thus reduce to two- or threefold application
of Eq.\ (\ref{overl2});\bibnote[laplace]{The situation is similar to the GTO basis, where
$\nabla^2 g(n_1,n_2,n_3)=g(n_1+2,n_2,n_3)+g(n_1,n_2+2,n_3)+g(n_1,n_2,n_3+2)$
reduces the kinetic energy integral to three overlap integrals within the same basis.}
\cite{Mitroy} common factors like $y_{lnn'}^{mm'}$ could be
reused.
(ii) The second direct method is based on the conversion of the Laplace operator
to a squared wave number in Fourier space:
\begin{eqnarray}
&&\int t_{n,m,s}(\alpha,{\bf r-A})\nabla^2_{\bf r}t_{n',m',s'}(\beta,{\bf r-B}) d^3r\nonumber \\
&=&
-\int \frac{d^3k}{(2\pi)^3}e^{-i{\bf k}\cdot{\bf C}}k^2 t_{n,m,s}(\alpha,{\bf k})
t_{n',m',s'}(\beta,-{\bf k}).
\end{eqnarray}
The common result of both approaches may be summarized as
\begin{eqnarray}
&&\int t_{n,m,s}(\alpha,{\bf r-A})\left(-\frac{\nabla^2_{\bf r}}{2}\right)
t_{n',m',s'}(\beta,{\bf r-B}) d^3r\nonumber \\
&=&
2^{n+n'+1}(4\pi)^2\sqrt{\frac{\pi N_{n,m}N_{n',m'}}{(2n+1)!!(2n'+1)!!\alpha^{s+3}\beta^{s'+3}}}
e^{-\gamma C^2}
\nonumber \\
&\times&
\gamma^{(n+n'+5)/2}
\Gamma\left(n+\frac{3}{2}+\frac{s}{2}\right)
\Gamma\left(n'+\frac{3}{2}+\frac{s'}{2}\right)
\nonumber \\
&\times&
\!\sum_{\sigma=0}^{s/2}\frac{(-s/2)_\sigma}{\sigma !\Gamma(n+\frac{3}{2}+\sigma)}
\left(\frac{\gamma}{\alpha}\right)^\sigma
\sum_{\sigma'=0}^{s'/2}\frac{(-s'/2)_{\sigma'}}{\sigma' !\Gamma(n'+\frac{3}{2}+\sigma')}
\left(\frac{\gamma}{\beta}\right)^{\sigma'}
\nonumber \\
&\times&
\sum_{l=|n-n'|}^{n+n'}
(-)^{(n-n'-l)/2}
(\gamma C^2)^{l/2}
\nonumber \\
&\times&
\Gamma\left(\frac{3}{2}+\frac{n+n'+l}{2}+\sigma+\sigma'+1\right)
y_{lnn'}^{mm'}(\hat{\bf C})
\nonumber \\
&\times&
\!\!\!\sum_{\tilde\sigma=0}^{(n+n'-l)/2+\sigma+\sigma'+1}
\frac{(\frac{l-n-n'}{2}-\sigma-\sigma'-1)_{\tilde\sigma}}{\tilde\sigma !
\Gamma(l+\frac{3}{2}+\tilde\sigma)}(\gamma C^2)^{\tilde \sigma},
\label{kinenr}
\end{eqnarray}
which is Eq.\ (\ref{overl2}) multiplied by $2\gamma$ plus the
replacement $\sigma+\sigma'
\to \sigma+\sigma'+1$ at three places in each of the last two lines of both equations.
The limit of common center ($C=0$) is
\begin{eqnarray}
&&\int t_{n,m,s}(\alpha,{\bf r-A})\left(-\frac{\nabla^2_{\bf r}}{2}\right)
t_{n',m',s'}(\beta,{\bf r-A}) d^3r\nonumber \\
&=& \delta_{n,n'}\delta_{m,m'}
\frac{2^{2n+4} \pi N_{n,m}}{(2n+1)!! \alpha^{(s+3)/2}\beta^{(s'+3)/2}}
\nonumber \\
&\times&
\gamma^{n+5/2}
\Gamma\left(n+\frac{3}{2}+\frac{s}{2}\right)
\Gamma\left(n+\frac{3}{2}+\frac{s'}{2}\right)
\nonumber \\
&\times&
\sum_{\sigma=0}^{s/2}\frac{(-s/2)_\sigma}{\sigma !\Gamma(n+\frac{3}{2}+\sigma)}
\left(\frac{\gamma}{\alpha}\right)^\sigma
\sum_{\sigma'=0}^{s'/2}\frac{(-s'/2)_{\sigma'}}{\sigma' !\Gamma(n+\frac{3}{2}+\sigma')}
\left(\frac{\gamma}{\beta}\right)^{\sigma'}
\nonumber \\
&\times&
\Gamma\left(n+\frac{5}{2}+\sigma+\sigma'\right) .
\end{eqnarray}

\subsection{1-Particle Coulomb Integral}\label{sec:Coul1}
The Fourier technique also suits the attack on the Coulomb integral:
\begin{eqnarray}
&&\int t_{n,m,s}(\alpha,{\bf r-A})\frac{1}{|{\bf r}-{\bf r'}|}t_{n',m',s'}(\beta,{\bf r-B}) d^3rd^3r'\nonumber \\
&=&
4\pi\int \frac{d^3k}{(2\pi)^3}\frac{e^{-i{\bf k}\cdot{\bf C}}}{k^2} t_{n,m,s}(\alpha,{\bf k})
t_{n',m',s'}(\beta,-{\bf k})
\nonumber \\
&=&
2^{n+n'}16\pi^3\sqrt{\frac{\pi N_{n,m}N_{n',m'}}{(2n+1)!!(2n'+1)!!\alpha^{s+3}\beta^{s'+3}}}
e^{-\gamma C^2}
\nonumber \\
&\times&
\gamma^{(n+n'+1)/2}
\Gamma\left(n+\frac{3}{2}+\frac{s}{2}\right)
\Gamma\left(n'+\frac{3}{2}+\frac{s'}{2}\right)
\nonumber \\
&\times&
\sum_{\sigma=0}^{s/2}\frac{(-s/2)_\sigma}{\sigma !\Gamma(n+\frac{3}{2}+\sigma)}\left(\frac{\gamma}{\alpha}\right)^\sigma
\sum_{\sigma'=0}^{s'/2}\frac{(-s'/2)_{\sigma'}}{\sigma' !\Gamma(n'+\frac{3}{2}+\sigma')}\left(\frac{\gamma}{\beta}\right)^{\sigma'}
\nonumber \\
&\times&
\!\!\!\sum_{l=|n-n'|}^{n+n'}
(-)^{(n-n'-l)/2}
(\gamma C^2)^{l/2}
\frac{\Gamma(\frac{3}{2}+\frac{n+n'+l}{2}+\sigma+\sigma'-1)}{\Gamma(l+\frac{3}{2})}
\nonumber \\
&\times&
y_{lnn'}^{mm'}(\hat{\bf C})
_1F_1( \frac{l-n-n'}{2}-\sigma-\sigma'+1;l+\frac{3}{2};\gamma C^2) .
\label{coulenr}
\end{eqnarray}
The limit of common center is
\begin{eqnarray}
&&\int t_{n,m,s}(\alpha,{\bf r-A})\frac{1}{|{\bf r}-{\bf r'}|}t_{n',m',s'}(\beta,{\bf r-A}) d^3rd^3r'\nonumber \\
&=& \delta_{n,n'}\delta_{m,m'}
\frac{2^{2n+3} \pi^2 N_{n,m}}{(2n+1)!! \alpha^{(s+3)/2}\beta^{(s'+3)/2}}
\nonumber \\
&\times&
\gamma^{n+1/2}
\Gamma\left(n+\frac{3}{2}+\frac{s}{2}\right)
\Gamma\left(n+\frac{3}{2}+\frac{s'}{2}\right)
\nonumber \\
&\times&
\sum_{\sigma=0}^{s/2}\frac{(-s/2)_\sigma}{\sigma !\Gamma(n+\frac{3}{2}+\sigma)}
\left(\frac{\gamma}{\alpha}\right)^\sigma
\sum_{\sigma'=0}^{s'/2}\frac{(-s'/2)_{\sigma'}}{\sigma' !\Gamma(n+\frac{3}{2}+\sigma')}
\left(\frac{\gamma}{\beta}\right)^{\sigma'}
\nonumber \\
&\times&
\Gamma\left(n+\frac{1}{2}+\sigma+\sigma'\right) .
\end{eqnarray}
In the general case of $l<n+n'$ or $\sigma+\sigma'>0$, the representation
\begin{eqnarray}
_1F_1( \frac{l-n-n'}{2}-\sigma-\sigma'+1;l+\frac{3}{2};\gamma C^2) && \nonumber \\
=\sum_{\tilde\sigma=0}^{(n+n'-l)/2+\sigma+\sigma'-1}
\frac{(\frac{l-n-n'}{2}-\sigma-\sigma'+1)_{\tilde\sigma}}{\tilde\sigma !
(l+\frac{3}{2})_{\tilde\sigma}}(\gamma C^2)^{\tilde \sigma} &&
\end{eqnarray}
lets Eq.\ (\ref{coulenr}) look like Eq.\ (\ref{overl2})
multiplied by $\pi/\gamma$
plus the replacement $\sigma+\sigma' \to \sigma+\sigma'-1$ at three places.
The special case of $l=n+n'=\sigma+\sigma'=0$ may be paraphrased with
Eq.\ (7.1.21) of Ref.\ \onlinecite{AS},
\begin{equation}
_1F_1(1;\frac{3}{2},\gamma C^2)=\sqrt{\frac{\pi}{\gamma C^2}}
\frac{e^{\gamma C^2}}{2}\mbox{erf}\sqrt{\gamma C^2}.
\label{F11erf}
\end{equation}
The remaining cases of $l=n+n'>0, \sigma+\sigma'=0$
could be recursively attached to the representation (\ref{F11erf})
with Eq.\ (13.4.6) of Ref.\ \onlinecite{AS},
\begin{equation}
_1F_1(1;l+\frac{3}{2},\gamma C^2)
=\frac{l+1/2}{\gamma C^2}\left\{
_1F_1(1;l+\frac{1}{2},\gamma C^2) -1
\right\}.
\end{equation}

Some basic observations on the numerical evaluation of Eqs.\ (\ref{overl2}),
(\ref{kinenr}) and (\ref{coulenr}) are:
\begin{itemize}
\item
$y_{lnn'}^{mm'}=0$ if $l+n+n'$ is odd.
Therefore each second term in the sum over $l$ may be skipped.
\item
All appearances of the $\Gamma$-function have the form
$\Gamma(j+3/2)=(2j+1)!!\sqrt{\pi}/2^{j+1}$ with small integers $j$,
and may be placed in a lookup table. (The associated factors $\sqrt{\pi}$
may be dropped, since the same number of $\Gamma$'s appears in numerators and
denominators.)
\item
In general, the number of terms in the sums is considerably smaller than
the number of GTO pairs in the ``dissociated'' products. The focus
shifts to efficient implementation of
the $\bar y_{lnn'}^{mm'}$ in Eqs.\ (\ref{firsty})--(\ref{lasty}).
\end{itemize}

\section{Table of Products}\label{sec:produ}
The 2-Particle Coulomb Integral
\begin{eqnarray}
\int && t_{n,m,s}(\alpha,{\bf r-A})t_{\bar n,\bar m,\bar s}(\bar \alpha,{\bf r'}-\bar {\bf A})
\frac{1}{|{\bf r}-{\bf r'}|} \nonumber \\
&\times& t_{\tilde n,\tilde m,\tilde s}(\tilde \beta,{\bf r'}-\tilde {\bf B})
t_{n',m',s'}(\beta,{\bf r-B})
d^3rd^3r'
\end{eqnarray}
may be computed by (i)
expansion of the product
$t_{n,m,s}(\alpha,{\bf r-A})t_{n',m',s'}(\beta,{\bf r-B})$
into a sum of terms
$\propto t_{\cdot,\cdot,\cdot}(\alpha+\beta,{\bf r}-{\bf P})$,
all centered at ${\bf P} \equiv (\alpha {\bf A}+\beta {\bf B})/(\alpha+\beta)$,
(ii)
expansion of the product
$t_{\bar n,\bar m,\bar s}(\bar \alpha,{\bf r'}-\bar {\bf A})
t_{\tilde n,\tilde m,\tilde s}(\tilde \beta,{\bf r'}-\tilde {\bf B})$
likewise into a sum of terms
$\propto t_{\cdot,\cdot,\cdot}(\bar\alpha+\tilde\beta,{\bf r}-\tilde{\bf P})$,
all centered at
$\tilde {\bf P}\equiv (\bar \alpha \bar{\bf A}+\tilde\beta \tilde {\bf B})/(\bar\alpha+\tilde\beta)$,
and 
(iii)
computing the remaining sum over 1-Particle Coulomb Integrals
with centers at ${\bf P}$ and $\tilde{\bf P}$
as proposed in Sec.\ \ref{sec:Coul1}\@.

To implement the first two steps, an expansion of a product
of two oscillator functions must be at hand.
Its evaluation is straight forward as proposed here, though tedious:
$t_{n,m,s}(\alpha,{\bf r}-{\bf A})$ and $t_{n',m',s'}(\beta,{\bf r}-{\bf B})$
are both expanded as described in App.\ \ref{appC} to obtain two sums
of HGTO's. They are merged with the product rule of HGTO's:
Those $g$ are shifted to the common center ${\bf P}$
\begin{eqnarray}
{\bf r}-{\bf A}&=& {\bf r}-{\bf P}+\frac{\beta}{\alpha+\beta}{\bf C} \\
{\bf r}-{\bf B}&=& {\bf r}-{\bf P}-\frac{\alpha}{\alpha+\beta}{\bf C}
\end{eqnarray}
where ${\bf C}={\bf B}-{\bf A}=(C_x,C_y,C_z)$.
Combined at ${\bf P}$, the product is a sum of $g$ of quantum
numbers up to $n+n'+s+s'$, and finally re-converted into a sum over
$t_{.,.,.}(\alpha+\beta,{\bf r}-{\bf P})$ as demonstrated in App.~\ref{appA}\@.
To simplify the intermediate steps, one actually uses CGTO's and the
associated
\begin{equation}
\tilde t_{n,m,s}\equiv r^s\tilde t_{n,m}=t_{n,m,s}/(2\alpha)^n
\end{equation}
to work this out. Switching from the $\tilde t$ representation
 to the $f$ representation is done with
the coefficients of the table in Sec.\ \ref{secmain}.
To switch back one builds a table with the projection technique of App.~\ref{appA}
that starts as follows:
\begin{eqnarray*}
& f(0,0,0)= \tilde t_{0,0,0} ; & \\
& f(0,0,1)= \tilde t_{1,0,0} ; & \\
& f(0,1,0)= \tilde t_{1,-1,0}; & \\
& f(1,0,0)= \tilde t_{1,1,0} ; & \\
& f(0,0,2)= \frac{1}{3}(\tilde t_{2,0,0}+\tilde t_{0,0,2}) ; & \\
& f(0,2,0)= -\frac{1}{6}\tilde t_{2,0,0}-\frac{1}{2}\tilde t_{2,2,0}
+\frac{1}{3}\tilde t_{0,0,2} ; & \\
& f(2,0,0)= -\frac{1}{6}\tilde t_{2,0,0}+\frac{1}{2}\tilde t_{2,2,0}
+\frac{1}{3}\tilde t_{0,0,2} ; & \\
& f(0,1,1)= \tilde t_{2,-1,0} ; & \\
& f(1,0,1)= \tilde t_{2,1,0} ; & \\
& f(1,1,0)= \tilde t_{2,-2,0} ; & \\
& f(0,0,3)= \frac{1}{5}\tilde t_{3,0,0}+\frac{3}{5}\tilde t_{1,0,2} ; & \\
& f(0,3,0)= -\frac{1}{4}\tilde t_{3,-3,0}-\frac{3}{20}\tilde t_{3,-1,0}
+\frac{3}{5}\tilde t_{1,-1,2} ; & \\
& f(0,1,2)= \frac{1}{5}(\tilde t_{3,-1,0}+\tilde t_{1,-1,2}) ; & \\
& f(1,0,2)= \frac{1}{5}(\tilde t_{3,1,0}+\tilde t_{1,1,2}) ; & \\
& f(0,2,1)= -\frac{1}{10}\tilde t_{3,0,0}-\frac{1}{2}\tilde t_{3,2,0}+\frac{1}{5}\tilde t_{1,0,2} ; & \\
& f(2,0,1)= -\frac{1}{10}\tilde t_{3,0,0}+\frac{1}{2}\tilde t_{3,2,0}+\frac{1}{5}\tilde t_{1,0,2} ; & \\
& f(1,2,0)= -\frac{1}{4}\tilde t_{3,3,0}-\frac{1}{20}\tilde t_{3,1,0}
+\frac{1}{5}\tilde t_{1,1,2} ; & \\
& f(2,1,0)= \frac{1}{4}\tilde t_{3,-3,0}-\frac{1}{20}\tilde t_{3,-1,0}
+\frac{1}{5}\tilde t_{1,-1,2} ; & \\
& f(1,1,1)= \tilde t_{3,-2,0} ; & \\
& f(0,0,4)= \frac{1}{35}\tilde t_{4,0,0}+\frac{1}{5}\tilde t_{0,0,4}+\frac{2}{7}\tilde t_{2,0,2} ; & \\
& f(2,2,0)=\frac{1}{280}\tilde t_{4,0,0}+\frac{1}{15}\tilde t_{0,0,4}
-\frac{1}{21}\tilde t_{2,0,2}-\frac{1}{8}\tilde t_{4,4,0}
. &
\end{eqnarray*}

Examples of products obtained with a Maple\cite{maple} implementation follow. The $\simeq$
signals the notational conventions that
the first $\tilde t$ on the left hand side has
the arguments $(\alpha,{\bf r}-{\bf A})$, that the second $\tilde t$
on the left hand side has the arguments $(\beta,{\bf r}-{\bf B})$,
that all $\tilde t$ on the right hand side have the arguments $(\alpha+\beta,{\bf r}-{\bf P})$,
and that the right hand side is to be multiplied by $\exp(-\gamma C^2)$.
The full writing of the first line would be
\begin{eqnarray}
&& \tilde t_{0,0,0}(\alpha,{\bf r}-{\bf A})
\tilde t_{0,0,0}(\beta,{\bf r}-{\bf B}) \nonumber \\
&& =\exp(-\gamma C^2)\tilde t_{0,0,0}
(\alpha+\beta,{\bf r}-{\bf P}).
\end{eqnarray}
Reduced exponentials $\alpha_r\equiv \alpha/(\alpha+\beta)$ and
$\beta_r\equiv \beta/(\alpha+\beta)$ are also inserted to simplify the notation.
\begin{eqnarray*}
 &&\tilde t_{0,0,0}\tilde t_{0,0,0}\simeq\tilde t_{0,0,0} ; \\
 &&\tilde t_{0,0,2}\tilde t_{0,0,0}\simeq \beta_r^2 C^2
\tilde t_{0,0,0}+\tilde t_{0,0,2} \\
&&\; +2\beta_r(C_x\tilde t_{1,1,0}
 +C_y\tilde t_{1,-1,0} +C_z\tilde t_{1,0,0}) ; \\
&& \tilde t_{0,0,2}\tilde t_{0,0,2}\simeq \alpha_r^2\beta_r^2 C^4
\tilde t_{0,0,0} +\tilde t_{0,0,4} \\
&&\; +(\alpha_r^2+\beta_r^2-\frac{4}{3}\alpha_r\beta_r)C^2\tilde t_{0,0,2} \\
&&\;
+2\alpha_r\beta_r(\alpha_r-\beta_r)C^2
(C_x\tilde t_{1,1,0}+C_y\tilde t_{1,-1,0}+C_z \tilde t_{1,0,0}) \\
&&\;
-2(\alpha_r-\beta_r)
(C_x\tilde t_{1,1,2}+C_y\tilde t_{1,-1,2}+C_z \tilde t_{1,0,2}) \\
&&\;
+\frac{2}{3}\alpha_r\beta_r (C^2-3C_z^2)\tilde t_{2,0,0} \\
&&\;
-8\alpha_r\beta_r
(C_xC_y\tilde t_{2,-2,0}+C_xC_z\tilde t_{2,1,0}+C_yC_z \tilde t_{2,-1,0}) \\
&&\;
-2\alpha_r\beta_r (C_x^2-C_y^2)\tilde t_{2,2,0}; \\
&&
\tilde t_{1,0,0}\tilde t_{0,0,0}\simeq\beta_r C_z\tilde t_{0,0,0}+\tilde t_{1,0,0} ; \\
&&
\tilde t_{1,1,0}\tilde t_{0,0,0}\simeq\beta_r C_x\tilde t_{0,0,0}+\tilde t_{1,1,0} ; \\
&&\tilde t_{1,-1,0}\tilde t_{0,0,0}\simeq\beta_r C_y\tilde t_{0,0,0}+\tilde t_{1,-1,0} ; \\
&&\tilde t_{1,-1,0}\tilde t_{0,0,2}\simeq\alpha_r^2\beta_r C_yC^2\tilde t_{0,0,0}
+(\beta_r-\frac{2}{3}\alpha_r)C_y\tilde t_{0,0,2} \\
&&\; +\alpha_r(\alpha_r C^2-2\beta_r C_y^2)\tilde t_{1,-1,0}
+\tilde t_{1,-1,2} \\
&&\;  -2\alpha_r\beta_r(C_xC_y\tilde t_{1,1,0}+C_yC_z\tilde t_{1,0,0}) \\
&&\; -2\alpha_r (C_x\tilde t_{2,-2,0} +C_z\tilde t_{2,-1,0}) \\
&&\; +\alpha_r C_y(\frac{1}{3}\tilde t_{2,0,0}+\tilde t_{2,2,0}) ; \\
&&\tilde t_{1,0,0}\tilde t_{0,0,2}\simeq\alpha_r^2\beta_r C_zC^2\tilde t_{0,0,0}
+(\beta_r-\frac{2}{3}\alpha_r)C_z\tilde t_{0,0,2} \\
&&\; +\alpha_r(\alpha_r C^2-2\beta_r C_z^2)\tilde t_{1,0,0}
+\tilde t_{1,0,2} \\
&&\;  -2\alpha_r\beta_r(C_xC_z\tilde t_{1,1,0}+C_yC_z\tilde t_{1,-1,0}) \\
&&\; -2\alpha_r (C_x\tilde t_{2,1,0} +C_y\tilde t_{2,-1,0}) -\frac{2}{3}\alpha_r C_z\tilde t_{2,0,0} ; \\
&&\tilde t_{1,-1,0}\tilde t_{1,-1,0}\simeq-\alpha_r\beta_rC_y^2\tilde t_{0,0,0}
+\frac{1}{3}\tilde t_{0,0,2} \\
&&\;  -(\alpha_r-\beta_r)C_y\tilde t_{1,-1,0} 
-\frac{1}{6}\tilde t_{2,0,0}-\frac{1}{2}\tilde t_{2,2,0} ; \\
&&\tilde t_{1,0,0}\tilde t_{1,-1,0}\simeq-\alpha_r\beta_rC_yC_z\tilde t_{0,0,0}
+\tilde t_{2,-1,0} \\
&&\; -\alpha_r C_y\tilde t_{1,0,0}+\beta_r C_z\tilde t_{1,-1,0} ; \\
&&\tilde t_{1,0,0}\tilde t_{1,0,0}\simeq-\alpha_r\beta_rC_z^2\tilde t_{0,0,0}
+\frac{1}{3}(\tilde t_{2,0,0}+\tilde t_{0,0,2}) \\
&&\;  -(\alpha_r-\beta_r)C_z\tilde t_{1,0,0}  ; \\
&&\tilde t_{1,1,0}\tilde t_{1,-1,0}\simeq-\alpha_r\beta_rC_xC_y\tilde t_{0,0,0}
+\tilde t_{2,-2,0} \\
&&\; -\alpha_r C_y\tilde t_{1,1,0}+\beta_r C_x\tilde t_{1,-1,0} ; \\
&&\tilde t_{1,1,0}\tilde t_{1,1,0}\simeq-\alpha_r\beta_rC_x^2\tilde t_{0,0,0}
+\frac{1}{3}\tilde t_{0,0,2} \\
&&\;  -(\alpha_r-\beta_r)C_x\tilde t_{1,1,0} 
-\frac{1}{6}\tilde t_{2,0,0}+\frac{1}{2}\tilde t_{2,2,0}
.
\end{eqnarray*}

\section{Absolute Norms}\label{secAbs}
 
If GTO's represent orbitals or wave functions, their squares represent particle
densities, and normalization follows from integrals like (\ref{normg})
or (\ref{normt}).
The first power rather than the second one specifies local densities, if
these functions are the constituents of density fitting functions.\cite{DunlapPRA42,JaffeJCP105,GohCPL264}
The measure
$\int g(n_1,n_2,n_3)d^3r=(\pi/\alpha )^{3/2}\delta_{n,0}$
is generally zero and does not provide a useful substitute to
Eqs.\ (\ref{normg}), (\ref{normt}) and (\ref{tnmnorm}) for that reason.
The absolute norm as calculated below is the next simple alternative.
It quantifies how many particles have to be moved from some region of
space to others to realize the specific density relocation, and becomes useful
to qualify the relative importance of terms with products of expansion
coefficients and the fitting functions.
 
The integrated absolute value of $g(n_1,n_2,n_3)$ is a product of
three integrals,
\begin{equation}
\int |g(n_1,n_2,n_3)| d^3r = G(n_1)G(n_2)G(n_3),
\label{gfactor}
\end{equation}
with
\begin{eqnarray}
G(n_j)&\equiv& \int_{-\infty}^{\infty} \alpha ^{n_j/2} |H_{n_j}(\sqrt \alpha  u)|e^{-\alpha u^2} du
\nonumber \\
&=&2\alpha ^{(n_j-1)/2}\int_0^{\infty}
|H_{n_j}(u)|e^{-u^2} du
.
\end{eqnarray}
$G(n_j)$ is a sum of $1+\lfloor n_j/2\rfloor$ integrals delimited by the positive roots
of $H_{n_j}$. Each integral is solved via Eq.\ (7.373.1) of Ref.\ \onlinecite{GR}
or Eq.\ (22.13.15) of Ref.\ \onlinecite{AS}:
\begin{eqnarray*}
G(0)&=&\sqrt{\pi/\alpha }\approx 1.772453850905516 \alpha ^{-1/2} ;\\
G(1)&=&2 ; \\
G(2)&=&4\sqrt{2\alpha /e}\approx 3.431055539842827 \alpha ^{1/2} ; \\
G(3)&=&4\alpha \left(1+4e^{-3/2}\right)\approx 7.570082562374877 \alpha  ; \\
G(4)&=&4(\alpha /e)^{3/2}\bigg[e^{-\sqrt{6}/2}H_3\left(\sqrt{\frac{3}{2}
+\frac{\sqrt 6}{3}}\right) \\
&&\quad -e^{\sqrt{6}/2}H_3\left(\sqrt{\frac{3}{2}-\frac{\sqrt 6}{3}}\right)
\bigg] \\
&\approx& 19.855739152211958 \alpha ^{3/2} ;\\
G(5)&\approx& 59.2575529009459587 \alpha ^2 ; \\
G(6)&\approx& 195.90006551027769 \alpha ^{5/2} ; \\
G(7)&\approx& 704.821503307929489499 \alpha ^3 .
\end{eqnarray*}
 
The integral $\int |f(n_1,n_2,n_3)|d^3r$ is calculated with ease
via Eqs.\ (3.461.2) and (3.461.3) of Ref.\ \onlinecite{GR}\@.
 
The linear combinations defined in Sec.\ \ref{secmain} may be evaluated in
spherical coordinates and decompose into products of integrals over $r$ [handled
by Eq.\ (3.461) of Ref.\ \onlinecite{GR}], $\varphi$ (yielding $2\pi$ or $4$
for $m=0$ or $m\neq 0$, respectively)
and $\theta$ [handled by determining the roots of $P_n^m(\cos\theta)$ and
decomposition into sub-intervals].
``Monomic'' cases like $t_{1,1}$, $t_{2,1}$ or $t_{3,-2}$, which relate
$t_{n,m}$ to a single HGTO, are
already represented by Eq.\ (\ref{gfactor}) and not listed again.
\begin{eqnarray*}
\int |t_{2,0}|d^3r & = &8\pi\sqrt{\frac{\pi}{3\alpha }}
        \approx 25.7190053432553290 \alpha ^{-1/2} ; \\
\int |t_{2,2}|d^3r & = & 8\sqrt{\pi/\alpha }
        \approx 14.1796308072441282 \alpha ^{-1/2} ; \\
\int |t_{3,0}|d^3r & = &  104\pi/5
        \approx 65.345127194667699360 ; \\
\int |t_{3,1}|d^3r & = &  \frac{224}{5}-16\arcsin(5^{-1/2})+4\pi \\
        &\approx& 49.94800887034627509; \\
\int |t_{3,2}|d^3r & = &  16 ; \\
\int |t_{3,3}|d^3r & = &  12\pi
        \approx 37.69911184307751886 ; \\
\int |t_{4,0}|d^3r & = & \frac{192}{245} \pi\sqrt{35\pi \alpha }\Big(
  (\sqrt{30}+3)\sqrt{15-2\sqrt{30}} \\
  &&\qquad +(\sqrt{30}-3)\sqrt{15+2\sqrt{30}}
  \Big) \\
 & \approx& 765.99700145937577804 \alpha ^{1/2}; \\
\int |t_{4,1}|d^3r & = &  8\sqrt{\pi \alpha }\left(1+128/7^{3/2}\right) \\
        &\approx& 112.180027342566280\alpha ^{1/2}; \\
\int |t_{4,2}|d^3r & = &  32\sqrt{\pi \alpha }\left(1+34/7^{3/2}\right) \\
        &\approx& 160.8439445477562992 \alpha ^{1/2}; \\
\int |t_{4,-2}|d^3r & = &  16\sqrt{\pi \alpha }\left(1+34/7^{3/2}\right) \\
        &\approx&  80.42197227387814961 \alpha ^{1/2}; \\
\int |t_{4,3}|d^3r & = &  24\sqrt{\pi \alpha }
        \approx 42.538892421732384655 \alpha ^{1/2}; \\
\int |t_{4,4}|d^3r & = &  64\sqrt{\pi \alpha }
        \approx 113.4370464579530257 \alpha ^{1/2}; \\
\int |t_{4,-4}|d^3r & = &  16\sqrt{\pi \alpha }
        \approx 28.3592616144882564 \alpha ^{1/2} ; \\
\int |t_{5,0}|d^3r & = & \frac{64}{567}\pi \left( 1701+640\sqrt{70}\right) \alpha \\
        &\approx& 2501.97074380428359 \alpha  ; \\
\int |t_{5,1}|d^3r & \approx & 594.462027151417576 \alpha  ; \\
\int |t_{5,2}|d^3r & = &  1024 \alpha /9= 113.\bar{7} \alpha  ; \\
\int |t_{5,-2}|d^3r & = &  512 \alpha /9=56.\bar{8} \alpha  ; \\
\int |t_{5,3}|d^3r & = &  \frac{8\alpha }{81}
        \left( 3064\sqrt{2}-2916\arcsin(1/3)+729\pi\right)\\
        &\approx& 556.287108307150907 \alpha ; \\
\int |t_{5,4}|d^3r & = &  128 \alpha  ; \\
\int |t_{5,-4}|d^3r & = &  32 \alpha  ; \\
\int |t_{5,5}|d^3r & = &  120\pi \alpha 
        \approx 376.991118430775189 \alpha 
.
\end{eqnarray*}
The absolute norms of the generalized oscillator functions read
\begin{equation}
\int |t_{n,m,s}(\alpha,{\bf r-A})|d^3r=
\alpha^{-s/2}\left(\frac{n+3}{2}\right)_{s/2}\int |t_{n,m}|d^3r
.
\end{equation}

\section{Summary}
A set of basis functions has been defined by recombination of HGTO's
or CGTO's
such that the members with common center are orthogonal and complete with
respect to the angular variables.
They turn out to be eigenfunctions of the isotropic
harmonic oscillator. A specific generalization of those allows (i) to keep
the computational advantage of sparse overlap and kinetic energy integrals,
(ii) backup by the tabulated GTO's in case of need for all integrals that
are multilinear in the orbitals
(2-particle Coulomb Integrals), still use of ``direct''
alternative formulas for overlap and kinetic energy
integrals, and (iii) to maintain the vector space of functions covered by GTO's.

\begin{acknowledgments}
The work was supported by the Quantum Theory Project at the
University of Florida and
grant DAA-H04-95-1-0326 from the U.S.A.R.O.
\end{acknowledgments}
 
\appendix
\section{Method of Expansion}\label{appA}
 
The representations in Sec.\ \ref{secmain} were obtained by re-engineering the straight-forward
solution to the inverse problem as follows:
\begin{enumerate}
\item
expanding all $(n+1)(n+2)/2$ GTO's of a fixed $n$ into
the complete set of $Y_l^m$ ($0\le l\le n$; $-l\le m\le l$).
For each  triple $(n_1,n_2,n_3)$,  this expansion implies the calculation of
$(n+1)^2$ integrals of the type
$\int  g(n_1,n_2,n_3)Y_l^{m*}\sin\theta d\theta d\varphi$.
Most of these vanish due to the selection rules
(i) $n_1+n_2+n_3+l$ even and
(ii) $l+m+n_3$ even.
One may also use that the expansion coefficient in front of
$Y_l^{-m}$ is the complex conjugate of the one in front of $Y_l^m$.
The pedestrian's way to calculate the integral over $\theta$
and $\varphi$ is (a) transformation of $g(n_1,n_2,n_3)$ into
spherical coordinates via Eq.\ (\ref{gdef}) and $x=r\cos\varphi\sin\theta$
etc. This results in a product of two integrals, one over $\theta$
and one over $\varphi$.
(b) The integral over $\varphi$ becomes
elementary if all occurrences of $\sin\varphi$ and $\cos\varphi$ are
substituted with Euler's formula.
(c) The substitution $\cos\theta=u$ reduces the integral over $\theta$
to a Hypergeometric Function $_3F_2$ (p.\ 183 of Ref.\ \onlinecite{BarnesQJPA39})\@.

\item
gathering and recombining all $Y_l^m$ within each expansion in terms of
$Y_l^{m,c}$ and $Y_l^{m,s}$.
 The list at $n\le 3$ (partial list at $n=4$) then reads:
 \begin{eqnarray*}
 &g(0,0,0)\simeq 2Y_0^{0,c} ; & \\
 &g(0,0,1)\simeq \frac{4}{\sqrt{3}} R Y_1^{0,c} ; & \\
 &g(1,0,0)\simeq \frac{4}{\sqrt{3}}R Y_1^{1,c} ; & \\
 &g(0,0,2)\simeq (\frac{8}{3} R^2-4)Y_0^{0,c}
 +\frac{16}{3\sqrt{5}} R^2 Y_2^{0,c} ; & \\
 &g(1,0,1)\simeq \frac{8}{\sqrt{15}} R^2 Y_2^{1,c} ; & \\
 &g(1,1,0)\simeq \frac{8}{\sqrt{15}} R^2 Y_2^{2,s} ; & \\
 &g(2,0,0)\simeq (\frac{8}{3} R^2-4)Y_0^{0,c}
 -\frac{8}{3\sqrt{5}} R^2 Y_2^{0,c}+\frac{8}{\sqrt{15}} R^2Y_2^{2,c} ; & \\
 &g(0,0,3)\simeq \frac{8}{5}\sqrt{3}R(2R^2-5)Y_1^{0,c}
 +\frac{32}{5\sqrt{7}} R^3 Y_3^{0,c} ; & \\
 &g(1,0,2)\simeq \frac{8}{5\sqrt{3}}R(2R^2-5)Y_1^{1,c}
 +\frac{32}{5}\sqrt{\frac{2}{21}} R^3 Y_3^{1,c} ; & \\
 &g(1,1,1)\simeq \frac{16}{\sqrt{105}}R^3 Y_3^{2,s} ; & \\
 &g(2,0,1)\simeq \frac{8}{5\sqrt{3}}R(2R^2-5)Y_1^{0,c}
 -\frac{16}{5\sqrt{7}} R^3 Y_3^{0,c} & \\
 &+\frac{16}{\sqrt{105}} R^3 Y_3^{2,c} ; & \\
 &g(2,1,0)\simeq \frac{8}{5\sqrt{3}}R(2R^2-5)Y_1^{1,s}
 -\frac{8}{5}\sqrt{\frac{2}{21}} R^3 Y_3^{1,s} & \\
 &+8\sqrt{\frac{2}{35}} R^3 Y_3^{3,s} ; & \\
 &g(3,0,0)\simeq \frac{8}{5}\sqrt{3}R(2R^2-5)Y_1^{1,c}
 -\frac{8}{5}\sqrt{\frac{6}{7}} R^3 Y_3^{1,c} & \\
 & +8\sqrt{\frac{2}{35}} R^3 Y_3^{3,c} ; & \\
 &g(0,0,4)\simeq ( \frac{32}{5}R^4-32R^2+24)Y_0^{0,c} & \\
  &+\frac{64}{7\sqrt{5}}R^2(2R^2-7) Y_2^{0,c}
 +\frac{256}{105} R^4 Y_4^{0,c} ; & \\
 &g(1,0,3)\simeq \frac{16}{7}\sqrt{\frac{3}{5}}(2R^2-7)Y_2^{1,c}
 +\frac{64}{21}\sqrt{\frac{2}{5}} R^4 Y_4^{1,c} ; & \\
 &g(1,1,2)\simeq \frac{16}{7\sqrt{15}}R^2(2R^2-7)Y_2^{2,s}
 +\frac{64}{21\sqrt{5}} R^4 Y_4^{2,s} ; & \\
 &g(2,0,2)\simeq (\frac{32}{15}R^4-\frac{32}{3}R^2+8)Y_0^{0,c} & \\
 &+\frac{16}{21\sqrt{5}}R^2(2R^2-7)Y_2^{0,c}
  +\frac{16}{7\sqrt{15}}R^2(2R^2-7)Y_2^{2,c} & \\
 &-\frac{128}{105} R^4 Y_4^{0,c}
 +\frac{64}{21\sqrt{5}} R^4 Y_4^{2,c} ; & \\
 &g(2,1,1)\simeq \frac{16}{7\sqrt{15}}R^2(2R^2-7)Y_2^{1,s}
 -\frac{16}{21}\sqrt{\frac{2}{5}}R^4 Y_4^{1,s} & \\
 & +\frac{16}{3}\sqrt{\frac{2}{35}} R^4 Y_4^{3,s} ; & \\
 &g(2,2,0)\simeq (\frac{32}{15}R^4-\frac{32}{3}R^2+8)Y_0^{0,c}
 +\frac{32}{105} R^4 Y_4^{0,c} & \\
 & -\frac{32}{21\sqrt{5}}R^2(2R^2-7)Y_2^{0,c}
 -\frac{32}{3\sqrt{35}} R^4 Y_4^{4,c};  & \\
 &g(3,0,1)\simeq \frac{16}{7}\sqrt{\frac{3}{5}}R^2(2R^2-7)Y_2^{1,c} & \\
  &-\frac{16}{7}\sqrt{\frac{2}{5}}R^4 Y_4^{1,c}
 +\frac{16}{3}\sqrt{\frac{2}{35}} R^4 Y_4^{3,c}
 . &
 \end{eqnarray*}
 The symbol $\simeq$ means a
 factor $e^{-\alpha r^2}\alpha^{n/2}\sqrt{\pi}$ has been omitted
 each time at the right hand side, and $R$ stands short for $\sqrt{\alpha}r$.
The coefficients of $g(n_2,n_1,n_3)$ are derived from those of
$g(n_1,n_2,n_3)$ by replacing
$Y_l^{m,c}$ with $(-)^{\lfloor m/2\rfloor}Y_l^{m,c}$ if $m$ is even,
$Y_l^{m,c}$ with $(-)^{\lfloor m/2\rfloor}Y_l^{m,s}$ and vice versa if $m$ is odd,
and $Y_l^{m,s}$ with $-(-)^{\lfloor m/2\rfloor}Y_l^{m,s}$ if $m$ is even.
\item
selecting the subset of equations that contain $Y_n^{m,c}$ for fixed
$m$ and $n$, yielding, say, $1\le q(m,n)\le(n+1)(n+2)/2$ equations.
[One may generally find more stringent upper bounds for $q$ by
inspection of the selection rule (ii) given above.]
The remaining task is to find $q(m,n)$ numbers such that the linear
combination of these $q$ equations by these numbers does not contain
any terms $Y_{l\neq n}^{m,c}$ or $Y_l^{m,s}$.
This means computing a $q$-dimensional basis vector of
a kernel of a
matrix that contains all the expansion coefficients prior
to those $Y$ terms that are to be eliminated.
Generally this matrix is non-square and $r$-dependent.
\item
normalizing this $q$-dimensional vector with some arbitrariness
such that its components are small integers and that
$t_{n,m}/[(\alpha r)^n\exp(-\alpha r^2)Y_n^{m,c}]$
are positive numbers. Those
$q$ components are
the expansion coefficients in front of the $g(n_1,n_2,n_3)$ of the table
for $m\ge 0$.
Counting terms shows $q(6,2)=8$, $q(6,0)=10$ and $q(3,2)=2$, for example.
Experience shows that --- up to this normalization factor ---  the expansions
are unique for $n\le 8$ at least,
i.e., the aforementioned kernel is one-dimensional.
\item
performing steps 3 to 4 for the $Y_n^{m,s}$ in an
equivalent manner by elimination of terms $\propto Y_{l\neq n}^{m,s}$
and $\propto Y_l^{m,c}$ to obtain $t_{n,-m}$.
\end{enumerate}
An additional shortcut exists once $t_{n,m}$ and $t_{n,-m}$ are known
for a specific ``anchorage'' $m$.
Application of the ladder operators
\begin{equation}
L_{\pm}\equiv L_x\pm iL_y
=i(y\partial_z-z\partial_y)\pm (x\partial_z-z\partial_x)
\end{equation}
of angular momentum quantum mechanics on $Y_n^m$ yields
$Y_n^{m\pm 1}$.
Decomposition of the two equations
\begin{equation}
L_{\pm}(t_{n,m}+it_{n,-m})\propto t_{n,m\pm1}+it_{n,-m\mp 1}
\end{equation}
into 4 real-valued equations yields a similar recursion for $t_{n,m}$.
Effectively one applies
\begin{eqnarray*}
\lefteqn{(y\partial_z-z\partial_y) g(n_1,n_2,n_3)} \nonumber \\
&&=n_3g(n_1,n_2+1,n_3-1)-n_2g(n_1,n_2-1,n_3+1) ; \\
\lefteqn{(x\partial_z-z\partial_x) g(n_1,n_2,n_3)} \nonumber \\
&&=n_3g(n_1+1,n_2,n_3-1)-n_1g(n_1-1,n_2,n_3+1)
\end{eqnarray*}
term by term to a pair of equally normalized $t_{n,m}$ and $t_{n,-m}$.

The quickest alternative to obtain the expansion is
developed in App.~\ref{appX} further down.
 
\section{Correspondence Between Cartesian and Hermite GTO's}\label{appB}
 
A Fourier Transform switches from Cartesian to Hermite GTO's and vice versa.\cite{KaijserAQC10,CharskyTCA93}
All HGTO's $g(n_1,n_2,n_3)$ at the right hand side of the equations
of Sec.\ \ref{secmain} transform into their associated CGTO's
$f(n_1,n_2,n_3)$ in ${\bf k}$-space,
\begin{equation}
\int e^{i{\bf k}\cdot{\bf r}}g(n_1,n_2,n_3)d^3r=
\left(\frac{\pi}{\alpha }\right)^{3/2}i^n e^{-k^2/(4\alpha )}\prod_{j=1}^3k_j^{n_j},
\end{equation}
and the components within the $t_{n,m}(r,\theta_r,\varphi_r,\alpha )$
at the left hand sides keep their angular
dependence,
\begin{eqnarray}
\lefteqn{\int e^{i{\bf k}\cdot{\bf r}}(\alpha r)^n e^{-\alpha r^2}Y_n^m(\theta_r,\varphi_r)d^3r}
\nonumber \\
&&=\left(\frac{\pi}{\alpha }\right)^{3/2}i^n e^{-k^2/(4\alpha )}
\left(\frac{k}{2}\right)^n Y_n^m(\theta_k,\varphi_k).
\label{fourrep}
\end{eqnarray}
Subsequent re-insertion of ${\bf r}$ for ${\bf k}$ leads to the expansion
of $Y_n^{m,c}$ or $Y_n^{m,s}$
in terms of CGTO's. Residual dangling factors depend on $n$, but not
on individual $n_j$, and are constant within each expansion, which allows
to stay with the coefficients as shown.
 
\section{Direct Expansion Formula}\label{appX}
The observation in App.~\ref{appB}
leads to a closed formula of the linear expansions of Sec.\ \ref{secmain},
because it suffices to expand $t_{n,m}$ and actually
$r^nY_n^{m\{c,s\}}(\theta,\varphi)e^{-\alpha r^2}$ in terms of $f(n_1,n_2,n_3)$.
An explicit writing of Eq.\ (\ref{plmdef}) is given by a re-formulation
of Eq.\ (8.812) in Ref.\ \onlinecite{GR},\citenote{plnsign}
\begin{eqnarray}
P_l^m(\cos\theta)&=&2^l\sin^{m}\theta \nonumber \\
&\times& \sum_{\nu=0}^{l-m}\left(\frac{m+\nu+1-l}{2}\right)_l\frac{\cos^{\nu}\theta}{\nu! (l-m-\nu)!}.
\label{thetaexp}
\end{eqnarray}
Terms with odd $l-m-\nu$ do not contribute to the sum (\ref{thetaexp}).
A synopsis of Eqs.\ (1.331.1--3) of Ref.\ \onlinecite{GR} is
\begin{equation}
\left.
\begin{array}{r} \cos(m\varphi) \\ \sin(m\varphi)
\end{array}
\right\}
=\sum_{j=\left\{ {0,2,4\ldots \atop 1,3,5\ldots}
\right.}^m (-)^{\lfloor j/2\rfloor}{m \choose j}\cos^{m-j}\varphi \sin^j \varphi .
\label{phiexp}
\end{equation}
The product of Eqs.\ (\ref{thetaexp}) and (\ref{phiexp}) is inserted into
Eqs.\ (\ref{Ylmc}) and (\ref{Ylms}). Simple counts of the powers of
the trigonometric functions constitute\citenote{Gaussbra}
\begin{eqnarray}
r^nY_n^{m\{c,s\}}(\widehat{\bf r})\propto&&\sum_{\nu=0}^{n-m}\left(\frac{m+\nu+1-n}{2}\right)_n
\frac{r^{n-\nu-m}}{\nu!(n-m-\nu)!} \nonumber\\
&&\times
\sum_{j=0(1)}^m(-)^{\lfloor j/2\rfloor}{m \choose j}x^{m-j}y^j z^{\nu}.
\end{eqnarray}
(The sum is over even $j$ for $Y_n^{m,c}$ and over odd $j$ for $Y_n^{m,s}$.)
The residual even power of $r$ is also broken down to Cartesian components
\begin{equation}
r^{n-\nu-m}=\sum_{\sigma_1,\sigma_2,\sigma_3\ge 0
\atop\sigma_1+\sigma_2+\sigma_3=(n-\nu-m)/2}
\frac{(\frac{n-\nu-m}{2})!}{\sigma_1!\sigma_2!\sigma_3!}
x^{2\sigma_1}y^{2\sigma_2}z^{2\sigma_3}.
\label{multinom}
\end{equation}
Whence
\begin{eqnarray*}
\tilde t_{n,m}&\propto&
\sum_{\nu=0}^{(n-|m|)/2}
\frac{(2n-2\nu-1)!! (-)^{\nu}}
{2^{\nu} (n-|m|-2\nu)!}
\nonumber \\
&\times& \sum_{\sigma_1,\sigma_2=0}^{\sigma_1+\sigma_2\le\nu}
\sum_{j=0(1)}^{|m|}
\frac{(-)^{\lfloor j/2\rfloor}}{\sigma_1!\sigma_2!(\nu-\sigma_1-\sigma_2)!}
{|m| \choose j}
\nonumber \\
&\times&
f(|m|-j+2\sigma_1,j+2\sigma_2,n-|m|-2(\sigma_1+\sigma_2)) .
\end{eqnarray*}
(The sum over $j$
is restricted to even $j$ if $m\ge 0$, and odd $j$ if $m<0$.)
Resummation enables performing the sum over $\nu$ analytically
to obtain Eq.\ (\ref{fasttnm2}).

\section{Expansion of Generalized Orbitals}\label{appC}

The expansion of $t_{n,m,s}=r^st_{n,m}$ in terms of HGTO's is obtained
as follows: $t_{n,m}$ is expanded as listed in Sec.\ \ref{secmain};
each $g(n_1,n_2,n_3)$ within the sum is written as a product according to Eq.\ (\ref{gdef})\@.
$r^s=(x^2+y^2+z^2)^{s/2}$ is expanded in Cartesian coordinates as in Eq.\ (\ref{multinom}),
and each of its Cartesian components recombined with the Hermite Polynomial of the same
Cartesian direction by repeated application of
\begin{equation}
2xH_{n_1}(x)=H_{n_1+1}(x)+2nH_{n_1-1}(x).
\end{equation}
$s=2$ just needs
\begin{equation}
x^2H_{n_1}(x)=\left\{
\begin{array}{ll}
\frac{1}{4}H_{n_1+2}(x)+(\frac{n_1}{2}+1)H_{n_1}(x) &  \\
\qquad +n_1(n_1-1)H_{n_1-2}(x),&(n_1\ge 2); \\
\frac{1}{4}H_3(x)+\frac{3}{2}H_1(x),&(n_1=1) ; \label{reexpand} \\
\frac{1}{4}H_2(x)+\frac{1}{2}H_0(x),&(n_1=0).
\end{array}
\right. \label{xH}
\end{equation}
and the equivalent expressions for $y^2H_{n_2}(y)$ and $z^2H_{n_3}(z)$.
The powers $s=4,6,\ldots$ follow by recursion of Eq.\ (\ref{xH}).

The case of CGTO's, $\tilde t_{n,m,s}\equiv r^s\tilde t_{n,m}$,
proceeds even simpler, because the obvious
$x^sf(n_1,n_2,n_3)=f(n_1+s,n_2,n_3)$ etc.\ does not need auxiliary formulas
like Eq.\ (\ref{reexpand}) to merge the components of $r^2$ with the
$f(n_1,n_2,n_3)$ contained in Sec.\ \ref{secmain}.
 
\section{Oscillator with Repulsive Core}\label{appD}

As Eq.\ (\ref{lapla2}) suggests, $t_{n,m,s}$ is an eigenfunction of the
spherical harmonic oscillator with superimposed repulsive core
potential
\begin{equation}
V({\bf r})=\frac{\hbar^2}{2M}(4\alpha^2 r^2+\frac{Q}{r^2}) .
\end{equation}
The standard discussion of the single-particle Schr\"odinger equation deals with a
fixed potential, hence fixed $Q$\@.
Its eigenfunctions $\psi_{nmj}$ are found with the method proposed
in Ref.\ \onlinecite{Messiah}:
\begin{equation}
\psi_{nmj}^{(p)}(\alpha,{\bf r})=e^{-\alpha r^2}
r^{p-1}{}_1F_1(-j;p+\frac{1}{2};2\alpha r^2)Y_n^{m\{c,s\}}(\hat{\bf r}),
\label{psinmdef}
\end{equation}
where
\begin{equation}
p\equiv  \frac{1}{2}+\sqrt{\frac{1}{4}+Q+n(n+1)},
\end{equation}
and $j=0,1,2,\ldots$ is the number of nodes in the radial function (the zero
at $r=0$ not counted). The $2n+1$-fold degenerate energy eigenvalues are
\begin{equation}
E_j=\frac{\hbar^2}{2M}4\alpha(j+p+\frac{1}{2}).
\end{equation}
[Additional ``accidental'' degeneracies occur if $p$ is integer, which includes
all cases of even, integer $Q=(q+n)(q-n-1)$ up to 200 for example.]
The Confluent Hypergeometric Functions may be written as Laguerre
Polynomials
\begin{equation}
_1F_1(-j;p+\frac{1}{2};2\alpha r^2)=\frac{j!}{(p+1/2)_{-j}}L_j^{(p-1/2)}(2\alpha r^2) ,
\end{equation}
which helps to derive the orthogonality relation with Eq.\ (7.414.3) of Ref.\ \onlinecite{GR},
\begin{equation}
\int \psi_{nmj}^{(p)}(\alpha,{\bf r})\psi_{n'm'j'}^{(p)}(\alpha,{\bf r})
d^3r
=\delta_{nn'}\delta_{mm'}\delta_{jj'}\frac{\Gamma(p+\frac{1}{2})}{2(2\alpha)^{p+1/2}}.
\label{psiorth}
\end{equation}
The $t_{n,m,s}$ defined in Eq.\ (\ref{tnmsdef}) refers to the case of integer
$Q=s(s+2n+1)$, integer $p=s+n+1$, and $j=0$. Since the factor $r^s(\alpha r)^n$
in (\ref{tnmsdef}) may be written as a linear combination of the polynomials
$r^{p-1}{}_1F_1()$ in Eq.\ (\ref{psinmdef}) and vice versa,
the set of basis functions $t_{n,m,s}(\alpha,{\bf r})$
spans the same phase space as the set of
$\psi_{nmj}^{(p)}(\alpha,{\bf r})$.

 A computational advantage of the eigenfunctions
(\ref{psinmdef}) is the orthogonality with respect to three indices manifested
in Eq.\ (\ref{psiorth}), whereas the $t_{n,m,s}$ are orthogonal with
respect to $n$ and $m$ but not with respect to $s$ [see Eq.\ (\ref{tortho})],
--- consequence of the fact that the $\psi_{nmj}$ are eigenfunctions of
a {\em fixed\/} potential and Hermite operator, whereas the various $t_{n,m,s}$ belong
to {\em different\/} potentials (different $Q$).
A disadvantage of the eigenfunctions (\ref{psinmdef}) is that their Fourier
representation has rather complicated sets of polynomials
which replace $_1F_1()$  in Eq.~(\ref{four2}), which takes negative effect on the
complexity of all formulas derived in  Sec.\ \ref{secIntgs}.

\bibliography{all}

\begin{thebibliography}{27}
\bibitem[{pln()}]{plnsign}
\bibinfo{note}{Definitions with an additional sign $(-)^m$ or $(-)^{(m+|m|)/2}$
  are also in frequent use.}
\bibitem[{Bib()}]{Bibnote3}
\bibinfo{note}{If orthogonality of the basis vectors is unimportant, other,
  much more general sets of linear combinations with fewer $g$-terms might help
  to reduce the total amount of time spent in numerical integrals calculations.
  (J. C. Boetter, priv.\ commun.)}.
\bibitem[{lap()}]{laplace}
\bibinfo{note}{The situation is similar to the GTO basis, where $\nabla ^2
  g(n_1,n_2,n_3)=g(n_1+2,n_2,n_3)+g(n_1,n_2+2,n_3)+g(n_1,n_2,n_3+2)$ reduces
  the kinetic energy integral to three overlap integrals within the same
  basis.}
\end{thebibliography}

\end{document}